\documentstyle[11pt,newpasp,twoside,epsf]{article}
\begin{document}

\title{Comparing Ultraviolet and H$\alpha$ Star Formation Rates}
\author{Eric F.~Bell and Robert C.~Kennicutt, Jr.}
\affil{Steward Observatory, 933 N. Cherry Ave., Tucson, AZ 85721, U.S.A.}

\begin{abstract}
We have used a sample of 43 star-forming galaxies imaged in the 
far-ultraviolet (FUV) by the UIT supplemented with 33 galaxies
observed by FAUST, to explore the consistency of UV and H$\alpha$ derived
star formation rates (SFRs).  We find, even before correction for
dust, that UV and H$\alpha$ SFRs are quantitatively 
consistent for low-luminosity galaxies, and
that higher luminosity galaxies have H$\alpha$ SFRs a factor of 1.5 higher than
their UV SFRs: this reflects the influence of dust.
Our results are consistent with a scenario where the 
UV dust extinction is a factor of $\la$2 larger than the H$\alpha$
dust extinction in a given galaxy, and that there is over a 4 magnitude
range of H$\alpha$ dust extinctions which correlate loosely with galaxy luminosity
such that low luminosity galaxies tend to have lower extinctions than
their higher luminosity counterparts.
\end{abstract}

\section{The Data}

We have selected a sample of 43 galaxies with FUV detections or
upper limits and H$\alpha$ luminosities from the sample of spiral galaxies 
imaged using the Ultraviolet Imaging Telescope (UIT).  
Magnitudes at an effective wavelength of 1567\,{\AA} 
were derived; from external comparisons, we estimate that our magnitudes
are accurate to at worst 0.4 mag, and typically 0.2 mag.
Magnitudes for a further 33 galaxies observed by FAUST (Deharveng 
et al. 1994) at 1650\,{\AA} were also included in the analysis as a check.  

H$\alpha$ luminosities are primarily taken from the literature, 
with the addition of some unpublished CCD data taken by the 
authors.  When appropriate, the total flux was corrected 
for $[$N{\,\sc ii}$]$ following Kennicutt (1983).
To determine the importance of dust, we also examined 
far-infrared (FIR) and thermal radio continuum luminosities.
The FUV to FIR ratio is a rough probe of dust attenuation.  
The H$\alpha$ to thermal radio ratio is a direct probe 
of the attenuation of the H$\alpha$ light as the two radiations are produced
by the same thermal electrons: the 
only disadvantage is that the thermal radio 
luminosity is difficult to measure accurately in most cases.

\begin{figure}
\hbox{\hspace{-0.7cm} \epsfysize=4.5cm\epsfbox{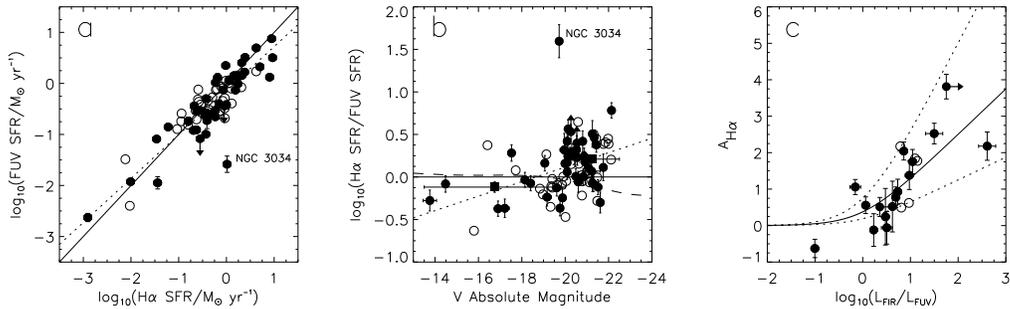} }
\caption{{\bf a)} Comparison of the UV and H$\alpha$ SFRs (calibrated using
Kennicutt 1998) for UIT ($\bullet$)
and FAUST ($\circ$) galaxies.  The solid line denotes equal SFRs, 
and the dotted line denotes the least squares fit to the data. 
{\bf b)} H$\alpha$ to UV SFR ratio with $V$ band absolute magnitude.
Symbols as in a): the dashed line denotes the trend in H$\alpha$
to FUV SFR ratio expected because of metallicity effects. 
{\bf c)} H$\alpha$ attenuation determined from the radio against 
the FIR to FUV luminosity ratio.  The solid line 
denotes the expected correlation 
if the H$\alpha$ attenuation is half of the FUV attenuation:
the dotted lines show the correlation if 
the H$\alpha$ attenuation is 1/4 of (lower line) or equal to (upper
line) the FUV attenuation.  }
\end{figure}

\section{Results}

In panels a and b of Fig.~1, we can see that the H$\alpha$
SFRs are a factor of $\sim$1.5 larger than the FUV SFRs for high luminosity
galaxies, and comparable for lower luminosity galaxies.  This
is consistent with a scenario in which both the H$\alpha$ and 
FUV radiation suffer attenuation (where the FUV attenuation 
is larger than the H$\alpha$ attenuation), and that the overall 
amount of attenuation varies as a function of galaxy luminosity
(e.g.\ Tully et al.\ 1998).

We explore the attenuation more directly in panel 
c, where we plot the H$\alpha$ attenuation determined from the 
H$\alpha$ to thermal radio ratio against the FIR to 
FUV luminosity ratio (a tracer of FUV attenuation, modulo
uncertainties from older stellar population heating of the dust).
The H$\alpha$ attenuations
are clearly correlated with the FUV attenuations, indicating
a real spread in overall spiral galaxy attenuation.  Furthermore, 
the data are consistent with a H$\alpha$ attenuation 
$\sim$ 1\,--\,2 times lower than the FUV attenuation (simple
models predict a factor of 4 offset: models that take into account
the different distribution of FUV and H$\alpha$ emitting sources
predict a factor of 2 offset).  A more thorough discussion of these
results is presented by Bell \& Kennicutt (2000).

\acknowledgements

This work was supported by NASA LTSA grant NAG5-8426 and NSF grant 
AST-9900789.

\end{document}